\documentclass{achemso}
\setkeys{acs}{articletitle = true}

\usepackage{amssymb,amsmath,graphicx,amsthm,wrapfig,epstopdf,esint,marginnote,afterpage,lipsum,color,tabularx,xcolor,soul}
\usepackage[normalem]{ulem}
\definecolor{myGreen}{rgb}{0,0.5,0}
\usepackage{graphicx}  
\usepackage{bm}
\usepackage{natbib}
\usepackage[version=4]{mhchem}
\usepackage{color,sidecap,fancyhdr}
\usepackage{hyperref}
\usepackage{cleveref,xcolor}

\newcommand{\bra}[1]{\left(#1\right)}
\newcommand{\Bra}[1]{\left[#1\right]}

\newcommand{\sss }{\sigma}

\title{From solvent free to dilute electrolytes: Essential components for a continuum theory}

\author{Nir Gavish}\affiliation{Department of Mathematics, Technion - IIT, Haifa, 3200003, Israel}

\author{Doron Elad}\affiliation{Department of Mathematics, Technion - IIT, Haifa, 3200003, Israel}

\author{Arik Yochelis}
\affiliation{Department of Solar Energy and Environmental Physics, Swiss Institute for Dryland Environmental and Energy Research, Blaustein Institutes for Desert Research, Ben-Gurion University of the Negev, Sede Boqer Campus, Midreshet Ben-Gurion 8499000, Israel}
\affiliation{Department of Physics, Ben-Gurion University of the Negev, Be'er Sheva 8410501, Israel}
\email{yochelis@bgu.ac.il}

\abbreviations{IL - ionic liquid, EDL - electric double layer,  PNP - Poisson-Nernst-Planck, SPNP - steric-PNP, MD - molecular dynamic, PB - Poisson-Boltzmann}

\begin{document}

\begin{abstract}
{The increasing number of experimental observations on highly concentrated electrolytes and ionic liquids show  qualitative features that are distinct from dilute or moderately concentrated electrolytes, such as self-assembly, multiple-time relaxation, and under-screening, which all impact the emergence of fluid/solid interfaces, and transport in these systems. Since these phenomena are not captured by existing mean field models of electrolytes, there is a paramount need for a continuum framework for highly concentrated electrolytes and ionic liquids. In this work, we present a self-consistent spatiotemporal framework for a ternary composition that comprises ions and solvent employing free energy that consists of short and long range interactions, together with a dissipation mechanism via Onsagers' relations. We show that the model can describe multiple bulk and interfacial morphologies at steady-state. Thus, the dynamic processes in the emergence of distinct morphologies become equally as important as the interactions that are specified in the equilibrium-free energy. The model equations not only provide insights to transport mechanisms beyond the Stokes-Einstein-Smoluchowski relations but also enables to qualitative recovery in the full range (three distinct regions) of non-monotonic electrical screening length that has been recently observed in experiments using organic solvent to dilute ionic liquids.}  
\end{abstract}

Concentrated electrolytes are being examined for a broad range of applications~\cite{Niedermeyer20127780,Perkin20125052,fedorov2014ionic,zhang2016nanoconfined}, and specifically for energy {related} devices, examples of which include dye sensitized solar cells, fuel cells, batteries and super-capacitors~\cite{Armand2009621,Wishart2009956,salanne2017ionic}. The optimized design of these devices requires the mechanistic spatiotemporal understanding of ionic arrangement and charge transport in electrolytes. Although the physicochemical aspects of electrolyte solutions have been extensively studied, a series of recent experimental and computational results~\cite{israelachvili1978measurement,attard1993asymptotic,leote1994decay,biesheuvel2007counterion,mansoori1971equilibrium,kirchner2015ion,Lee2015,smith2016electrostatic,Yochelis2015,makino2011charging,limmer2015interfacial,adar2017bjerrum} reflects knowledge gaps even in the context of basic science~\cite{Bazant:2009fp,iglivc2015nanostructures,Lee2015159,hayes2015structure,fedorov2014ionic,Yochelis2015,goodwin2017underscreening}. A robust spatiotemporal framework is thus required to advance electrochemically-based industrial applications of highly-concentrated electrolytes, e.g., fuel cells $\sim$1M (mol/liter) and batteries $\sim$10M.

Spatiotemporal theoretical formulation for electrolytes goes back to 1890's where mean-field Poisson-Nernst-Planck (PNP) framework was originated~\cite{nernst1889elektromotorische,planck1890ueber}.
The PNP model describes ions in electrolytes as isolated point charges which obey drift-diffusion transport under an electric potential, and indeed serves a basis for dilute electrolytes~\cite{Bard1980}. This description is valid for dilute electrolytes (below $\sim0.01$M), but is oversimplified for concentrated electrolytes where, for example, the finite size of the ions is important~\cite{Bikerman1942,booth1951dielectric}.  Consequently, extensive studies during the past centennial have led to major modifications of the mean-field approach to tackle the asymptotic electrical double layers (EDL) or electrokinetic properties, see representative reviews~\cite{varela2003exact,Bazant:2009fp,iglivc2015nanostructures,goodwin2017underscreening} and references therein.
Among the broad family of modified PNP and Poisson-Boltzmann (PB) models, we highlight for our purposes the steric-PNP (SPNP) approach that has been employed for studying permeability and selectivity in ion channels~\cite{horng2012pnp}, where ion concentrations exceed 10M. This approach accounts for Lennard-Jones type interactions between the ions, which are similar to the interactions employed in molecular dynamic (MD) simulations.

Theoretically (i.e., excluding the ion-pairing question~\cite{Lee2015159}) the upper limit of `concentrated electrolytes' pertains to molten salts, which about room temperature are also being referred to as ionic liquids (IL)~\cite{Armand2009621,Wishart2009956}, i.e., solvent-free electrolytes.
Besides the multiple applications, IL either with or without dilution also pose several physicochemical properties that extend the phenomenology of concentrated electrolytes: (\textit{i}) multiple timescales~\cite{makino2011charging,camci2016xps}, (\textit{ii}) self-assembly~\cite{Atkin20084164,celso2017direct,pontoni2017self}, and (\textit{iii}) non-monotonic variation (exhibiting qualitatively three distinct regions) of EDL with concentration~\cite{smith2016electrostatic,smith2017switching}, a.k.a., the phenomenon of {\em under-screening}. These phenomenologies emphasize that besides the vast progress in concentrated electrolytes, a unified framework that combines altogether the above mentioned phenomena is still missing~\cite{goodwin2017underscreening}.


Here we present a dynamical and thermodynamically consistent, unified continuum framework for \textit{ternary} media based on SPNP and Onsager's relations. Our goal is to present a theoretical framework by outlining the {essential features that are designed to pave the road toward a realistic description for concentrated electrolytes}: (\textit{i}) explicit densities of electrically positive, negative and neutral subsets (aqueous or organic carriers), (\textit{ii}) finite-size (steric) effects, and (\textit{iii}) selective energy dissipation to reflect mass transport.  While we follow the SPNP formulation, the starting point stems from IL~\cite{gavish2016theory,bier2017mean} {and introduce explicitly the solvent density}. 
The model {is qualitatively tested and appear to capture and relate, in broad parameter regimes, the above mentioned phenomena to effects of polarization~\cite{fedorov2014ionic} and self-assembly~\cite{hayes2015structure} to explain the origins of non-monotonic electrical screening length variation~\cite{smith2016electrostatic,smith2017switching}. In particular, at high concentrations, {the system is expected to indeed form different morphologies in both the EDL and bulk region} that are strongly dictated by spatiotemporal instabilities that drive the self-assembly process in the bulk, as in models for pure ionic liquids~\cite{gavish2016theory,lazaridis2017fluctuating} and charged polymer diblock copolymers~\cite{gavish2017spatially}.}

\section{Model equations for a ternary composition}

In general, PNP-type models consider the solvent as an effective background medium. This assumption, however, breaks in the high concentration limit, e.g., in ionic liquids. Respectively, we derive a ternary model by considering first model equations for asymmetric pure IL~\cite{bier2017mean} and introducing the solvent explicitly. Specifically, we assume a monovalent electrolyte comprising cations ($p$), anions ($n$) and solvent ($s$), identified by their the molar concentrations and that maintain uniform densities in terms of partial volume that is occupied by each species: 
\begin{equation}\label{eq:uniformDensity}
\frac{p}{p_{\max}}+\frac{n}{n_{\max}}+\frac{s}{s_{\max}}=1,
\end{equation}
where $p_{\max}$,~$n_{\max}$ and~$s_{\max}$ being the maximal concentrations (packing densities) of the cations, anions and solvent molecules, respectively.
{Following Onsagers' framework, cf.~\cite{gavish2016theory}}, the transport equations are given by 
\begin{subequations}\label{eq:dimensionalTernary}
\begin{equation} \label{disMech}
\frac{\partial}{\partial t}\left(\begin{array}{l}
p\\n\\s
\end{array}\right)= \nabla\cdot 
\,\begin{bmatrix}
p_{\max}-p & -p & - p \\
-n & n_{\max} -n &  -n \\
- s & - s & s_{\max}- s
\end{bmatrix}
\left(\begin{array}{l}
p\, \mathbf{J}^p\\ 
n\, \mathbf{J}^n\\
s\, \mathbf{J}^s
\end{array} \right).
\end{equation}
and the Coulombic interactions obey the standard Poisson's equation
\begin{equation}\label{eq:model_poss}
	\nabla \bra{\epsilon\nabla \phi}=q(n-p),
\end{equation}
\end{subequations}
where $q$ is the elementary charge and $\epsilon$ is the dielectric permittivity (assumed here to be constant). The respective fluxes are obtained from variation (functional derivative) of the free energy
\[
\mathbf{J}^{p}=\frac{\mu_B}{\overline{c}_{\max}} \nabla \dfrac{\delta \mathcal{F}}{\delta p}, \quad \mathbf{J}^{n}=\frac{\mu_B}{\overline{c}_{\max}} \nabla \dfrac{\delta \mathcal{F}}{\delta n}, \quad \mathbf{J}^{s}=\frac{\mu_B}{\overline{c}_{\max}} \nabla \dfrac{\delta \mathcal{F}}{\delta s},
\] 
where
\begin{eqnarray}
\nonumber \mathcal{F}&=& \int_{\Omega } \mathrm{d{\bf{x}}} \overbrace{k_B T \Bra{p \ln \frac{p}{\overline{c} } + n \ln \frac{n}{\overline{c}} + s \ln \frac{s}{\overline{s}}}}^{\text{entropy}} + \overbrace{\Bra{ q(p-n)\phi - \frac{\epsilon }{2} |\nabla \phi |^2}}^{\text{electrostatic}} \\
&&
+\underbrace{\bar{c}_{\rm max} \Bra{\frac{\beta }{n_{\max} p_{\max}}np + \mathcal{E}_0 \frac{\kappa ^2 }{2} \left( \left|\nabla \frac{p}{p_{\max}}\right|^2 +\left| \nabla \frac{n}{n_{\max}}\right|^2  \right)}}_{\text{local approximation of  Lennard-Jones interactions}},
\end{eqnarray}
$\Omega$ is the considered physical domain, $\mu_B$ is the mobility coefficient, $\beta$ is the interaction parameter (a.k.a. Flory parameter in the context of mixing energy for polymers) for the anion/cation mixture, $\mathcal{E}_0 \kappa ^2$ is the gradient energy coefficient with units of energy for the former and units of length for the latter, and $\bar{c}$ is the average concentration of ionic species and~$\bar{s}$ is the average concentration of solvent molecules, 
\[
\bar{c}:=\frac{1}{|\Omega|}\int_{\Omega} p\,\mathrm{d{\bf{x}}} = \frac{1}{|\Omega|}\int_{\Omega} n\,\mathrm{d{\bf{x}}},\quad \bar{s}:=\frac{1}{|\Omega|}\int_{\Omega}s\,\mathrm{d{\bf{x}}}.
\]
Note that global charge neutrality and the equal valence of anion and cations ensure that the average concentration of cations equals the average concentration of anion. This implies, together with the electrolyte incompressibility assumption~$\eqref{eq:uniformDensity}$, that the maximal average concentration (achieved when~$\bar{s}=0$), equals the harmonic average of~$p_{\max}$ and~$n_{\max}$ 
\[
\frac{1}{\bar{c}_{\max}}=\frac{1}{p_{\max}}+\frac{1}{n_{\max}}.
\]

\section{Model interpretation} 

To uncover the physical properties the key parameters of the model equations~\eqref{eq:dimensionalTernary}, we rewrite the equations in their dimensionless form (see SI for details) and for simplicity present only the equation for cations while the equation for anions is given in the SI:
	\begin{eqnarray}\label{eq:model}
		\label{eq:model_p}
		\frac{\partial p}{\partial t }&=&
		\nabla\cdot \bra{\mathbf{J}^p_{\rm PNP}+c\mathbf{J}^p_1+c^2 \mathbf{J}^p_2+c^3 \mathbf{J}^p_3},
	\end{eqnarray}
where
\begin{subequations}\label{eq:flux_p}
\begin{eqnarray}
	\label{eq:PNP_p}
	\mathbf{J}^p_{\mathrm{PNP}}&=&\underbrace{\nabla p+p\nabla\phi}_{\text{drift-diffusion}},\\
	\label{eq:J1_p}
	\mathbf{J}^p_1&=&\underbrace{\frac{2\Upsilon}{1+\gamma}p \nabla \bra{p+n}}_{\text{solvent entropy}}- \underbrace{p\mathbf{J}_{PNP}^p}_{\text{saturated mobility}} 
- \underbrace{p\mathbf{J}_{PNP}^n
	+
		\frac{\chi }{1+\gamma} p\nabla n}_{\text{inter-diffusion}}, \\
	\label{eq:J2_p}
	\mathbf{J}^p_2&=&-\underbrace{\frac{1}{1+\gamma}p\left[\sigma \nabla^3 p+\chi\nabla \bra{p  n}\right]}_{\text{steric effects}},\\
	\label{eq:J3_p}	
	\mathbf{J}^p_3&=&\underbrace{\frac{\sigma}{ \bra{1+\gamma}}p\left( p\nabla^3 p+n\nabla^3 n\right)}_{\text{steric effects}}.
	\end{eqnarray}
\end{subequations}
The normalized global ion concentration~$0\leq c\leq 1$ is a key parameter that relates between the ions and the solvent via~$\bra{1-c}s=1- c\bra{p+n}$. In particular, $c$ is scaled such that~$c=0$ corresponds to absence of ions (dilute limit), while~$c=1$ to absence of solvent (ionic liquid), i.e., powers of $c$ describe equal contributions of terms in the equations and do not imply a perturbative expansion. Here, $\sigma$ is a measure for the ratio between relative strength of short range steric and long range Coloumbic interactions, $\chi$ is also related to steric interactions rescaled to the ambient temperature,~$\gamma$ is the size ratio between cations and anions, and~$\Upsilon$ is the complementary size asymmetry between the solvent and ions. In particular,~$\Upsilon \sim 1$ for solvent molecules whose size is similar to the size of the ions (e.g., DMF), and $\Upsilon \gg 1$ for solvent molecules which are much smaller than the ions (e.g., water).

It is instructive to consider the ionic flux term-by-term to reveal the impact of different mechanisms on electrolyte behaviour as concentration is varied. For simplicity of the exposition we do so by choosing $\mathbf{J}^p$ while components of $\mathbf{J}^n$ are given in the SI.
At the dilute limit,~$c\ll1$, steric effects are perturbative, i.e.,~$\mathbf{J}^p\approx \mathbf{J}_{PNP}^p$.  Therefore, equation~\eqref{eq:model_p} reduces to the classical PNP theory.  
Similarly, in the solvent-free limit,~$c\simeq 1$, the model~\eqref{eq:model_p} reduces to the model for pure ionic liquids~\cite{bier2017mean} {that was shown to exhibit multiple timescale dynamics in the context of transient currents~\cite{gavish2016theory}}. To understand how electrolyte behaviour varies at intermediate concentrations, let us consider initially, terms of~$\mathbf{J}_1^p$, that is, $c$ contributions.  
The first term in $\mathbf{J}_1^p$ is related to \textit{solvent entropy} that reflects the solvent tendency to diffuse to region of high ionic concentration (thus, low solvent concentration). This effect becomes dominant when~$\Upsilon\gg1$, i.e., when solvent molecules are significantly smaller than the ions, since entropy per volume of many small particles is larger than the entropy of fewer larger particles. The second term corresponds to \textit{saturated mobility}, which gradually diminishes the transport of cations to cation-rich regions. This term prevents the formation of regions with a local concentration of constitutes beyond their packing density. The final two terms correspond to \textit{inter-diffusion} (a.k.a. mutual- or cross-diffusion) where species movement results in opposite directions of the electrochemical potential of anions and cations, respectively. Here, the motion reflects the increase/decrease in the local anion or cation concentrations which is partially balanced by decrease/increase of the total charge concentration, as well as a change in solvent concentration. 
Significantly, these mechanisms are not introduced `by-hand' to the model, they are results of the explicit inclusion of solvent and to a lesser extend the assumption of electrolyte incompressibility. Indeed, while Nernst-Planck dynamics arises from a random walk of isolated ions in an effective medium leading to Stokes-Einstein-Smoluchowski relations, the inclusion of solvent implies that species are not isolated. Thus, the movement of a cation, for example, involves pushing aside solvent molecules and possibly anions in his path and the movement of solvent to the void left behind him.

At higher orders, the combined effect of all terms can be appreciated in one space dimension, by neglecting the boundary conditions and considering the case $\gamma=\Upsilon=1$. In this case, substituting~\eqref{eq:model_poss} into~\eqref{eq:J1_p}, the flux takes the form of high order corrections for the drift $\mathbf{J}_1^p=p\nabla^3 \phi$, and thus, $\nabla \cdot \mathbf{J}^{p}_{1}$ gives rise to a fourth order derivative of the electric potential that addresses short range correlations~\cite{santangelo2006computing,liu2013correlated,Bazant:2011ha,blossey2017structural} (we note that originally the high order operator was introduced phenomenologically through a displacement field in~\eqref{eq:model_poss}). In the next section, we show that, under appropriate conditions, these short-range interactions give rise to self-assembled nano-patterns near charged interfaces and/or in the bulk.

\section{Analysis: Non-monotonic screening length}

\begin{figure}[tp]
\includegraphics[width=0.45\textwidth]{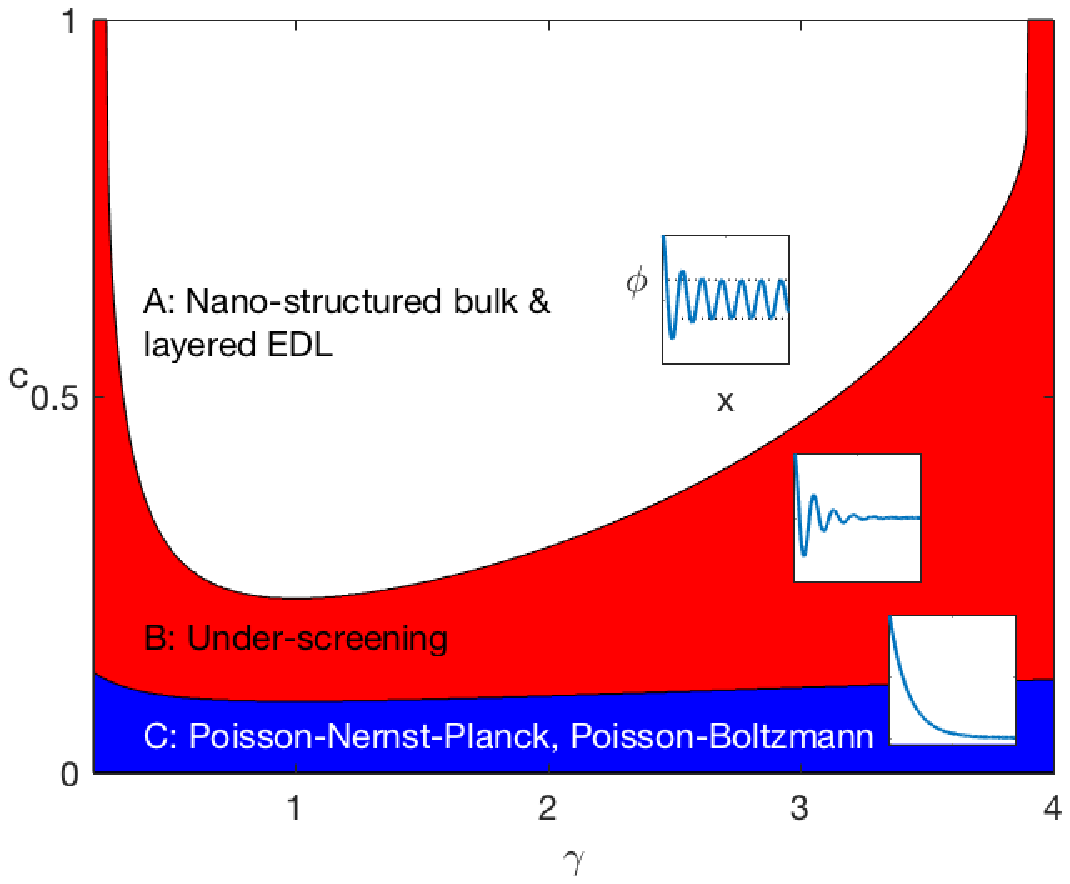}
	\caption{Parameter space spanned by concentration ($c$) and ion size asymmetry ($\gamma$), showing the transitions from monotonic EDL as described by the PNP or Poisson-Boltzmann model ({dark/blue shaded region}), region of oscillatory EDL giving rise to under-screening ({light/red shaded region}), and region of self-assembled bulk.  Parameters: $\Upsilon=4$, $\sigma=10$, $\chi=30$.} \label{fig:phaseDiagram}
\end{figure}

The broad family of generalized PNP models describes electrolyte solutions with a spatially uniform bulk~\cite{varela2003exact,Bazant:2009fp,EJM:10386443}. Eqs.~\ref{eq:model} gives rise to a richer picture of bulk and interfacial behavior, that is consistent with the picture arising in recent experimental and numerical studies~\cite{iglivc2015nanostructures,hayes2015structure,Atkin20084164,mezger2015solid,cui2016influence,maggs2016general,pontoni2017self}.
Indeed, using spatial linearization methods, see SI for details, we map the qualitatively distinct bulk and interfacial behaviors, and the corresponding regions in the parameter space spanned by ion size asymmetry~$\gamma$ and normalized concentration~$c$, see Figure~\ref{fig:phaseDiagram}. The results have been obtained in one space dimension ($\nabla \to \partial_x)$ and with no-flux boundary conditions (inert electrodes) for ions and Dirichlet for the potential (constant applied voltage): 
$J^p(x=\pm L/2)=0$, $\phi(x=\pm L/2)=\pm V/2$,
where $L$ is the physical domain size and $V$ is the applied voltage. 

{\em Region A}, describes the emergent self-assembly of bulk nano-structure, i.e., a spatially oscillatory profile throughout the whole domain, see top inset in Figure~\ref{fig:phaseDiagram}. Here, the region is obtained from the onset of the finite wavenumber instability~\cite{bier2017mean}, see SI for details. This region is characterized by highly concentrated electrolyte with weak ion size asymmetries (typical size rations of roughly 1:2 or 1:3). Indeed, the first observation for self-assembly was for protic IL with short alkyl groups~\cite{Atkin20084164}, although the phenomenon of self-assembly has a much broader context~\cite{hayes2015structure}. 

{\em Region C}, corresponds to a homogeneous bulk with a monotone electrical diffuse layer (EDL) structure, as shown in the bottom inset of Figure~\ref{fig:phaseDiagram}. This region is characterized by low enough concentrations so that steric effects are perturbative, and the model reduces to the PNP theory of dilute electrolytes (or Poisson-Boltzmann (PB) theory if steady states are of interest).  The latter also demonstrates that sufficiently diluted IL, either by solvent~\cite{mezger2015solid} {or possibly by} formation of ion-pairs~\cite{Gebbie2013,adar2017bjerrum}, can behave as dilute electrolytes. 

The intermediate {\em Region B}, describes a homogeneous bulk with an oscillatory potential and charge density profiles near a charged interface (a.k.a, a spatially oscillatory EDL), as shown in the middle inset. The onset of this region is computed using spatial dynamics methods~\cite{Yochelis2014,Yochelis2014b}, as detailed in the SI.  Notably, these decaying oscillations have also been referred to as \textit{over-screening}~\cite{Bazant:2011ha,levin2002electrostatic}. 

\begin{figure*}[tp] 
	\includegraphics[width=0.75\textwidth]{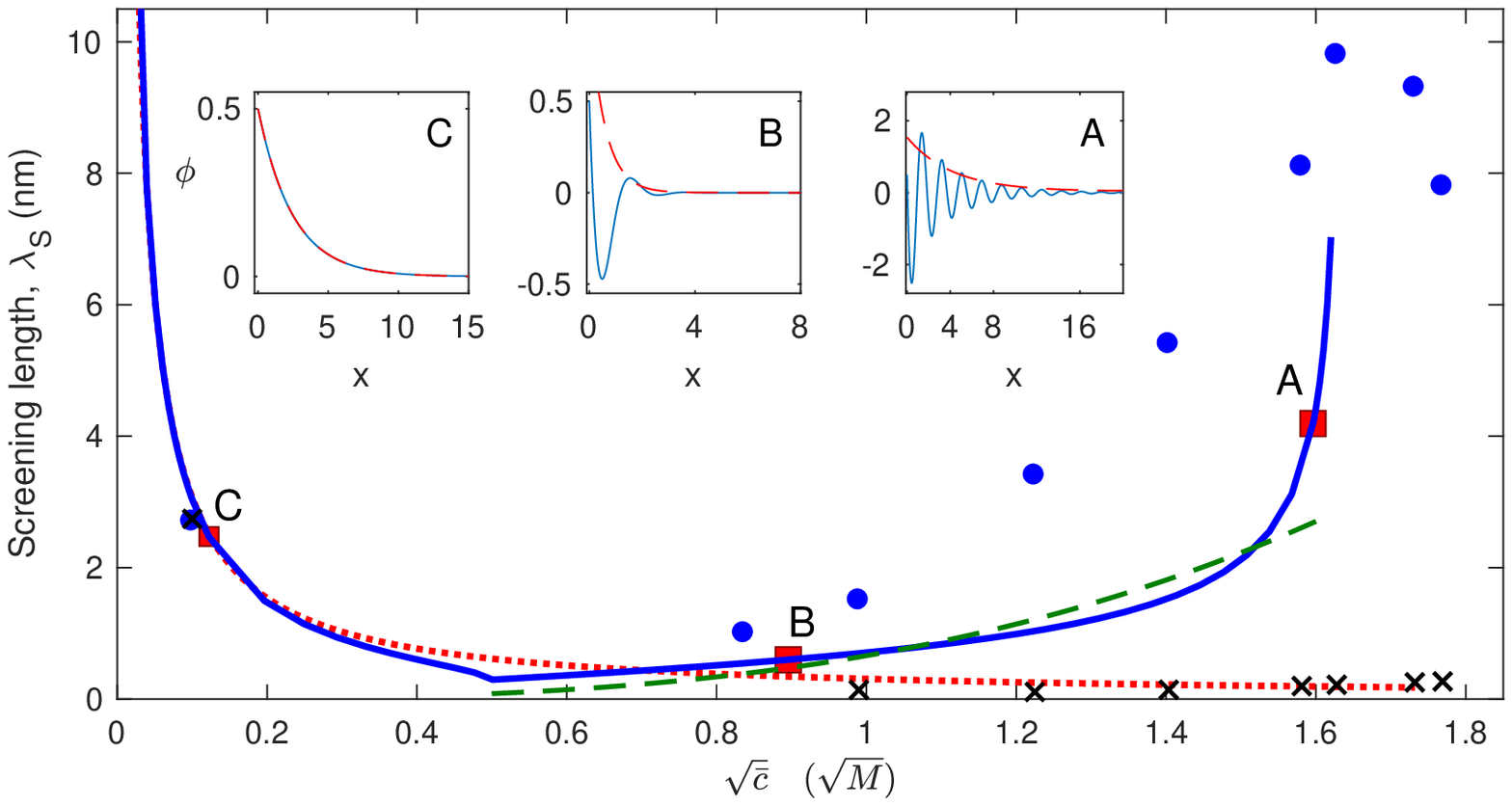}
	\caption{Screening-length as a function of concentration~$\bar{c}$ for ionic liquid, [\ce{C4C1Pyrr}][\ce{NTf2}], diluted with propylene carbonate (\ce{C4H6O3}); {the concentration is scaled according to dependence of Debye length}. Experimental data from~\cite{smith2016electrostatic} ($\bullet$) is superimposed with computed screening length (solid curve) for dimensional parameters that correspond to this ionic liquid, see SI for details. For comparison, we also plotted the monotonically decaying Debye-H\"{u}ckel approximation, $\lambda_D$  (dotted red curve), {the hypernetted chain correction~\cite{attard1993asymptotic} ($\times$),} and the curve~$2/3\bar{c}^{3/2}$  (dashed curve) to {allow comparison} with the recently presented scaling argument~\cite{lee2017scaling}.  The insets A-C present spatial profiles of the electric potential~$\phi$ at different concentrations (solid curve), see ($\blacksquare$), and the exponential envelop (dashed curve) that is computed by the real part of the spatial eigenvalues~\eqref{eq:eigs}, which is being used as a measure for the screening length. Dimensionless parameters:~$\gamma=1$, $\Upsilon=1$, $\chi=28$, $\sigma=35$,~$V=1$ and $L=200$.} \label{fig:RTIL}
\end{figure*}

The three regions in Fig.~\ref{fig:phaseDiagram}, imply that, 
as an ionic liquid or concentrated electrolyte solution with relatively weak ion-size asymmetry~($\gamma\approx 1$), is diluted, it exhibits a transition from nano-structured bulk and layered EDL at high ionic concentration (i.e., at region A) to layered EDL with homogeneous bulk (region B) and finally at low concentrations to a monotonic EDL structure.  The insets A-C in Figure~\ref{fig:RTIL} present steady-state solutions of~\eqref{eq:model} with parameters corresponding to the ionic liquid, [\ce{C4C1Pyrr}][\ce{NTf2}], diluted with propylene carbonate (\ce{C4H6O3}). Indeed, a transition from layered to monotonic EDL has been observed using high-energy x-ray reflectivity~\cite{mezger2015solid} and \textit{in situ} AFM~\cite{cui2016influence} within the range of parameters considered here. The molecular origin for the transition from monotonic to spatially oscillatory EDL, a.k.a. Kirkwood line, have been outlined using hypernetted chain~\cite{attard1993asymptotic} and mean spherical~\cite{Buzzeo20041106} approximations already in the early 1990s. {However, as has been outlined by Smith \textit{et al.}~\cite{smith2016electrostatic} and also demonstrated in Fig.~\ref{fig:RTIL}, there are significant differences in comparison with empirical measurements.}  Moreover, the transition from {region B to region A} appears to be unexplored by available mean-field models.  In what follows, we show that this transition is associated with an unexpected decrease in the screening length with increasing concentration~\cite{smith2016electrostatic,smith2017switching}, {see region~$\sqrt{\bar{c}}\gtrsim 1.6\sqrt{M}$ in Fig.~\ref{fig:RTIL}}. 


To examine the transition from region {C to B}, we use the size symmetric $\gamma=1$ case to derive analytic relations, which otherwise are obtained numerically, as shown in the SI. Above the monotonic to spatially oscillatory decay onset, the spatial decay is no longer being described by the PB approach and instead the screening length is better related to the envelope of the decaying oscillations~\cite{attard1993asymptotic}. In other words, the screening length is related to the spatial scale at which the bulk electroneutrality is violated~\cite{Yochelis2014,Yochelis2014b} ({\em tail} of oscillation amplitudes) rather than to the electrolyte behavior in the vicinity of the solid surface. Indeed, in Fig.~\ref{fig:RTIL}A, we observe that the envelope of the oscillatory tail (dashed curve) well describes the screening length, but does not describe the electrolyte behavior near the boundary, i.e., in the first two left spatial oscillations near $x=0$.

Following the re-definition of the screening length with the scale at which electroneutrality is departed, the formal computation involves linearization in space about a uniform bulk~$p\equiv p_0(c)$ and~$n\equiv n_0(c)$ corresponding to the normalized concentration~$c$,
\[
p=p_0+ \delta_p e^{\mu x},\quad n=n_0+ \delta_n e^{\mu x},\quad \phi = \delta_\phi e^{\mu x},
\]
where, $|\delta_p|,|\delta_n|,|\delta_\phi| \ll 1$ are the respective eigenvector components at the onset. In this case, the screening length is the reciprocal of the real part of the spatial eigenvalues,~$\lambda_S \sim {\rm Re}(1/\mu)$, that are associated with the eigenvalue problem, see SI for details. Particularly, in the size-symmetric case ($\gamma=\Upsilon=1$) explicit expressions of these eigenvalues can be derived, where out of the several eigenvalues, the relevant ones are given by
\begin{equation}\label{eq:eigs}
\mu^2_{\pm}=\frac{\chi c-4}{2c^2 \sigma}\left[  1   \pm \sqrt{ 1 -16 \sss c^2 / \left(4-\chi c \right)^2}  \right].
\end{equation}
Figure~\ref{fig:RTIL} presents the computed screening length as a function of concentration for parameters corresponding to a specific ionic liquid, [\ce{C4C1Pyrr}][\ce{NTf2}], diluted with propylene carbonate. In accordance with experimental observations, at low concentrations where the EDL structure monotonically decreases, the screening length identifies with the Debye length. As electrolyte concentration is increased above the Kirkwood point, $c_c= 4(\chi-4 \sqrt{\sigma})/\bra{\chi^2-16 \sigma} \simeq 0.077$, the envelope tail length increases and exceeds the Debye length. Respectively, in dimensional units the transition is at $0.48$ $\sqrt{M}$, which corresponds to the minimum in the IL screening length (Fig.~\ref{fig:RTIL}). Accordingly, increase in the {concentration} leads to the phenomenon of {under-screening}~\cite{smith2016electrostatic,smith2017switching}. {As introduced by~Lee \textit{et al}.~\cite{lee2017scaling}, {in a regime of concentration beyond the Kirkwood point}, we observe that the screening length roughly scales as~$c^{3/2}$ with respect to the Debye screening length}.

{While the transition from regions C and B can be formulated using near equilibrium methods, it is not true for the transition from region B to A. Here, the significance of spatiotemporal approach~\eqref{eq:dimensionalTernary} in the context of screening length structure becomes fundamental as the temporal bulk instability is approached, i.e., region A.} At the instability onset, the real part of the {relevant} eigenvalues~${\rm Re}(\mu_{\pm})\to 0$, and hence the screening length in a pure 1D system should extend to infinity, $\lambda_S \to \infty$. This onset is in fact {an analogue of the (temporal) finite wavenumber instability bifurcation point, as discussed in more detailed by Gavish \textit{et al.}~\cite{gavish2016theory,gavish2017spatially}. However, in reality the system is even beyond the 1D representation, so that secondary time-dependent zigzag instabilities~\cite{Cross1993851} of the bulk self-assembly {become dominant in an appropriate parameter regime}~\cite{gavish2016theory,gavish2017spatially}}. In this case, spatial oscillations of screening length are destroyed and the screening length shows a decreasing feature, cf.~\cite{lazaridis2017fluctuating}. Indeed, the second decay in screening length have been reported at high concentrations, see cf.~\cite{smith2016electrostatic,smith2017switching} and Figure~\ref{fig:RTIL}. {The phenomenological origin of this behavior requires thus, a spatiotemporal framework as opposed to methods which focus on equilibrium behavior in regions B and C~\cite{attard1996electrolytes,varela2003exact,Storey2012,goodwin2017underscreening}.}   

Consequently, as summarized in Fig.~\ref{fig:RTIL}, the above results qualitatively capture the experimental observation of all three regions of the non-monotonic screening length dependence on concentration, as well as quantitatively agree with the PB theory~\cite{smith2016electrostatic}. We have also tested consistency of our results with parameters that correspond to aqueous \ce{NaCl}, and obtained again a good qualitative comparison, as shown in Fig.~1 in the SI. 

\section{Conclusions} 

In this work, we have introduced a spatiotemporal thermodynamically consistent continuum mean-field framework for electrolytes that ranges from solvent free to dilute concentrations with explicit treatment of the solvent medium and finite size (steric) effects of both ion and solvent components. These two components introduce, by consistency, an additional transport mechanism of inter-diffusion and bulk nano-structuring by self-assembly. We conjecture that these are the {\em essential} features that any mean-field for electrolytes should comprise to provide a comprehensive description at a wide range of concentrations that approach a solvent-free case. However, in specific cases other approaches can be also efficient, for example, steric effects could be also introduced by using different approaches: a (repulsive) Yukawa potential, the hard sphere potential in Rosenfeld’s density functional theory of fluids~\cite{varela2003exact}, and asymptotic PB-type models~\cite{varela2003exact,BenYaakov:2011ek,giera2015electric,maggs2016general,adar2017bjerrum,blossey2017structural}. To emphasize that, we chose to focus on the screening length behavior at a full range of concentrations, which is beyond the available theories as it captures the effect of secondary transverse instabilities that explain the experimentally observed second decay at high concentrations~\cite{smith2016electrostatic,smith2017switching}. Notably, the advantage of spatiotemporal framework is in cases where understanding of temporal and/or {morphological bulk effects is required, such as self-assembly. A detailed analysis of secondary instabilities is beyond the scope of this letter and will be presented elsewhere}.



{The advantage of the developed framework is not only consistency with different previous mean-field approaches~\cite{Bikerman1942,attard1993asymptotic,Borukhov1997435,psaltis2011comparing,horng2012pnp,Storey2012,gavish2016theory,lazaridis2017fluctuating} but it} can be readily be extended to account for additional effects, e.g., ion-solvent interactions, ion-pairing~\cite{Marcus20064585,merlet2010internal,Hollóczki201416880,Lee2015159,adar2017bjerrum}, or other electrostatic corrections~\cite{schlumpberger2017simple}. These extensions should be advanced toward quantitative comparisons with experimental observations, beyond the qualitative trends depicted in see Fig.~\ref{fig:RTIL}. Consequently, we expect that current spatiotemporal framework will stimulate experiments that combine temporal methods with structural analysis~\cite{reichert2017molecular}, for example, transport properties such as multiple timescale relaxations~\cite{gavish2016theory} {that have been already shown for pure ionic liquids due to inter-diffusion and saturated mobility~\cite{gavish2016theory}}, have not been addressed here in detail, as systematic (for comparison purposes) experimental data is limited at large~\cite{makino2011charging,camci2016xps}.



\begin{acknowledgement}
This research was done in the framework of the Grand Technion Energy Program (GTEP) and of the BGU Energy Initiative Program, and supported by the Adelis Foundation for renewable energy research. N.G. acknowledges the support from the Technion VPR fund and from EU Marie--Curie CIG Grant 2018620.
\end{acknowledgement}


\bibliographystyle{achemso}
\providecommand{\latin}[1]{#1}
\providecommand*\mcitethebibliography{\thebibliography}
\csname @ifundefined\endcsname{endmcitethebibliography}
{\let\endmcitethebibliography\endthebibliography}{}

\end{document}